\documentclass[twocolumn, amssymb, amsmath, aps, prb, showpacs, 10pt]{revtex4-1}
\usepackage[dvips]{graphicx}
\usepackage[dvips]{color}
\begin{document}
\title{Nature of the glassy magnetic state in Cu$_{2.84}$Mn$_{0.44}$Al$_{0.72}$ shape memory alloy}
\author{S. Chatterjee$^1$}
\email{souvik@alpha.iuc.res.in} 
\author{S. Chattopadhyay$^2$}
\author{S. Giri$^2$}
\author{S. Majumdar$^2$}
\affiliation{$^1$UGC-DAE Consortium for Scientific Research, Kolkata Centre, Sector III, LB-8, Salt Lake, Kolkata 700 098, India}
\affiliation{$^2$Department of Solid State Physics, Indian Association for the Cultivation of Science, 2 A \& B Raja S. C. Mullick Road, Jadavpur, Kolkata 700 032, India}
\begin{abstract}
The magnetic ground state of the ferromagnetic shape memory alloy of nominal composition Cu$_{2.84}$Mn$_{0.44}$Al$_{0.72}$ was investigated. The sample shows reentry of a glassy magnetic phase below the martensitic transition temperature, which is found to have complex character with two distinct anomalies in the temperature dependent ac susceptibility data. The sample retains its glassy phase even below the second transition as evident from the magnetic memory measurements in different protocols. Existence of  two transitions along with their observed nature suggest that the system can be described by the mean field Heisenberg model of reentrant spin glass as proposed by Gabay and Toulous.~\cite{rsg-GT1} The sample provides a fascinating example where a Gabay-Toulous type spin glass state is triggered by a first order magneto-structural transition.

\end{abstract}
\maketitle

\section{Introduction}
The non-diffusive thermoelastic  Martensitic Transition (MT) in Ferromagnetic shape memory alloy (FSMA) is a notable example of magneto-structural instability among metallic alloys. FSMAs are bi-ferroic materials which combine ferroelasticity and ferromagnetism through the MT, and this magnetoelastic coupling helps to manipulate the ferroelastic domains of FSMA by magnetic field leading to numerous functional properties.~\cite{Ullakko-apl,Pons-msea} The electronic and magnetic ground states of FSMAs are often found to be quite intriguing due to the complex interplay between these two ferroic properties. 

\par
The present paper deals with a Cu-Mn-Al based FSMA with particular focus on the magnetic ground state of the sample. Localized nature of Mn moment is responsible for the magnetism in this system~\cite{Pardo-am}, and the magnetic coupling is mediated through Rudermann-Kittel-Kasuya-Yosida (RKKY) type exchange interaction between Mn atoms.~\cite{Webster-book}
Stoichiometric Cu$_2$MnAl is ferromagnetic in nature although its elements are all non-ferromagnetic.~\cite{Tajima-jpsj}
However,  Cu$_2$MnAl does not undergo martensitic type structural instability. Structural MT in Cu-Mn-Al alloys occurs far from the Heusler stoichiometric region.~\cite{Pardo-am, winkler-JMMM, prado-cumnal, kainuma} For compositions close to Cu$_{3-x}$Mn$_x$Al (0 $ < x <$ 1), the system generally crystallizes in a BCC $\beta$ phase at high temperature ($\sim$ 850$^{\circ}$ C). But with  proper heat treatment, cubic L2$_1$ phase can be stabilized above room temperature.~\cite{manosa-cumnal} On further cooling, cubic L2$_1$ phase undergoes MT to a more closed packed phase 18R type monoclinic structure.~\cite{prado-cumnal, wang-cumnal}

\begin{figure}[t]
\centering
\includegraphics[width = 8 cm]{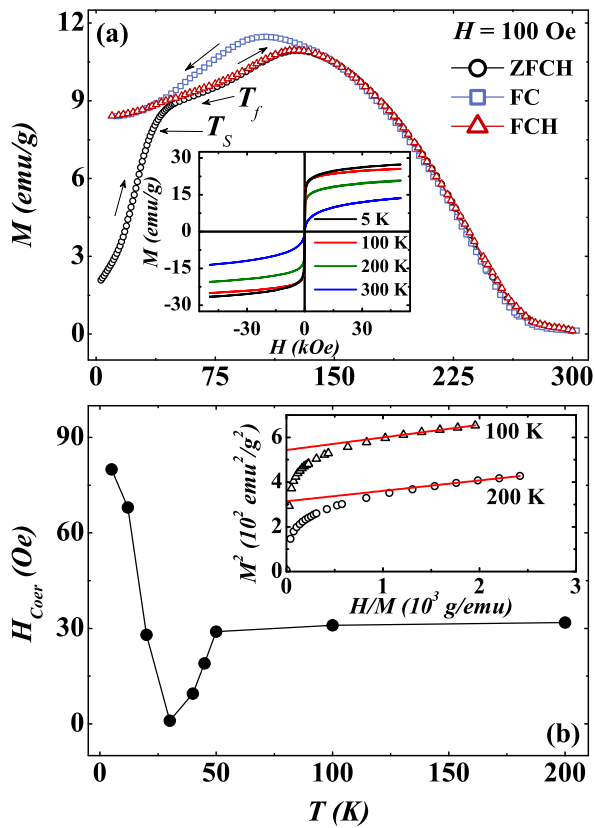}
\caption{(Color online) (a)  shows the variation of magnetization ($M$) as a function of temperature ($T$) in zero-field-cooled heating (ZFCH), field-cooling (FC) and field-cooled heating (FCH) protocol in presence of 100 Oe of applied magnetic field ($H$). Arrows indicate the direction of $T$ change. The inset of (a) shows the isothermal magnetization recorded at different constant temperatures.  The main panel of (b) shows the temperature variation of coercive field of the sample. The inset of (b) shows Arrot  ($M^2$ versus $H/M$) plot. }
\label{mtmh}
\end{figure}

\par
True magnetic nature of Cu-Mn-Al system of alloys is still not thoroughly understood. The stoichiometric Heusler compound Cu$_2$MnAl undergoes long range ferromagnetic (FM) order below 630 K.~\cite{obra} For the off-stoichiometric Cu-Mn-Al alloys undergoing thermoelastic MT, the magnetic state is quite complex. In literature there are reports on the observation of superparamagnetism and mictomagnetism below the MT.~\cite{yef,Pardo-am} The off-stoichiometry alloys with random site occupancy can have both FM and antiferromagnetic (AFM) correlations  leading to spin frustration. In fact  spin glass state is reported for Cu-Mn-Al alloys in low Mn concentration, which  turns into a superparamagnetic like state with increasing Mn.~\cite{obrado-sg, manosa-eng} Superparamagnetism is characteristically different from a spin glass or FM state, as in the former case the superspin clusters are mutually noninteracting.~\cite{beda} The reported metastability in Cu-Mn-Al alloys warrant further investigations to unveil the true nature of the glassy ground state. It is particularly important to address the origin of low-$T$  spin glass phase in Cu-Mn-Al, and  to what extent it is intrinsic to the sample and corresponds to those of conventional spin glasses. Keeping all these in mind, we chose the alloy Cu$_{2.84}$Mn$_{0.44}$Al$_{0.72}$  for the thorough characterization of its magnetic ground state. It can be thought of derived from Cu$_{3-x}$Mn$_x$Al ($x$ = 0.16) along with some excess Mn (0.28) being doped at the expanses of Al. Being high in Mn concentration, the sample undergoes long range FM ordering rather than being a superparamagnet. The ground state is found to be more complex with  spin glass like transition well below the region of MT.

\begin{figure}[t]
\centering
\includegraphics[width = 8 cm]{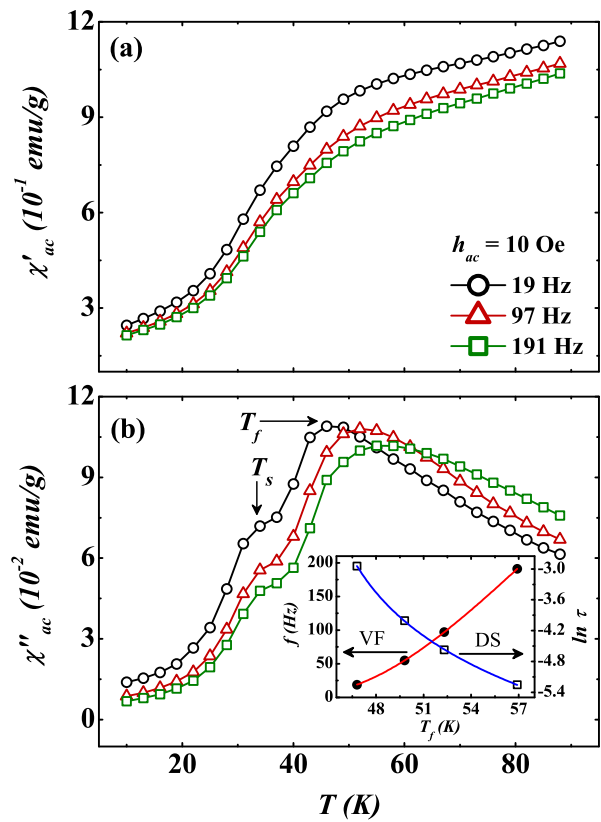}
\caption{(Color online) (a) Real and (b) imaginary parts of ac susceptibility as a function of temperature measured at three different applied frequencies of the ac signal. During measurement dc magnetic field was kept zero, whereas, magnitude of the applied ac field was 10 Oe. The inset of (b) shows the frequency dependence of peak in the imaginary part of ac susceptibility near $T_f$. Solid line (red) indicates the Vogel-Fulcher (VF) fit to the data. Relaxation time ($\tau$) is   plotted as a function of $T_f$ in the inset. Solid line (blue) indicates fitting in accordance with dynamical scaling (DS) model to the experimental data (see text for details).}
\label{acchi}
\end{figure}

\section{Experimental details}
The polycrystalline sample of nominal composition Cu$_{2.84}$Mn$_{0.44}$Al$_{0.72}$ was prepared by argon arc melting of the constituent elements. The ingot was homogenized at 800$^{\circ}$C for 20 minutes followed by a rapid quenching in ice water, which helps to stabilize the desired L2$_1$ phase at room temperature.~\cite{obrado-sg,manosa-eng} The sample was characterized by x-ray powder diffraction using Cu K$_{\alpha}$ radiation at room temperature. The crystal structure is found to be cubic L2$_1$-type with lattice parameter $a$ = 6.039 \AA. The dc magnetization ($M$) of the sample was measured on  Quantum Design SQUID magnetometer (MPMS, Ever-cool model). The ac susceptibility ($\chi_{ac}$) was measured on a commercial cryogen free low temperature system (Cryogenic Ltd., UK) using mutual inductance bridge technique.

\section{Results}
Fig.~\ref{mtmh} (a) depicts the temperature ($T$) variation of $M$ recorded in zero-field-cooled heating (ZFCH), field cooling (FC) and field-cooled heating (FCH) protocols in presence of 100 Oe of applied dc magnetic field ($H$). While cooling from 300 K, $M$ shows a sudden upturn around $T_C$ = 270 K which signifies the development of an FM-like state. The signature of MT is visible below about 150 K where FC and FCH data show thermal hysteresis. Separation between FCH and ZFCH data starts to emerge below $T_f$ = 65 K. However, significantly large irreversibility is only observed below $T_s$ = 35 K. Similar  fall in $M$ below the MT  was earlier observed in several Ni-Mn-Z (Z = Sb, Sn, In) alloys~\cite{nali1, nali2, zl} and it was  found to be related to the onset of glassy magnetic state and exchange bias effect.~\cite{sc-prb3, nali1, nali2}  

\par
We  measured isothermal $M$  as a function of $H$ at different  $T$ [see inset of fig.~\ref{mtmh} (a)]. The isotherms show typical FM like behavior with  an initial rise at low field followed by a tendency for saturation at higher fields. However, the $M(H)$ isotherm recorded  at 5 K does not show complete saturation. This indicates that either the magnetic anisotropy of the FM phase is very high or there is phase separation with some other magnetic phase coexisting with the majority FM fraction. All the isotherms show small but finite coercivity ($H_{coer}$) at least below $T_C$. Extrapolation of high field part of  Arrot plot ($M^2$ {\it vs.} $H/M$)~\cite{arrot} shows finite intercept on the vertical axis below $T_C$. This signifies non-zero spontaneous magnetization expected for a ferromagnetically ordered sample.  Fig.~\ref{mtmh} (b) shows the variation of $H_{coer}$ with $T$. At high temperature (down to 50 K), the observed coercivity is small and almost independent of $T$.  $H_{coer}$ starts to decrease below 50 K and it goes through a minimum at around 30 K with its value almost turning to zero. On further cooling, $H_{coer}$ rises and attains a value of 80 Oe at 5 K. This minimum point matches closely with the temperature $T_s$  which signifies the onset point for large thermomagnetic irreversibility.

\begin{figure}[t]
\centering
\includegraphics[width = 8 cm]{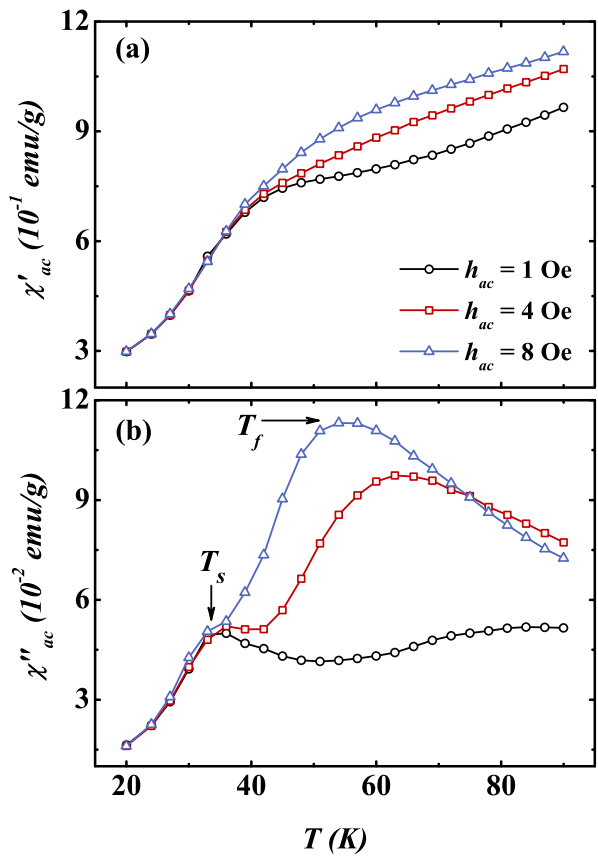}
\caption{(Color online) (a) and (b) depict temperature variation of real and imaginary part of ac susceptibility at three different amplitudes of the applied ac signal respectively. During measurement dc magnetic field and frequency of applied ac signal were kept constant at 0 Oe and 97 Hz respectively.}
\label{chiampl}
\end{figure}

\par
Evidently, it becomes important to know the significance of the temperatures $T_f$ and $T_s$. We performed ac susceptibility measurement at different constant frequencies to ascertain the presence of any spin-glass-like state in the studied alloy. Real ($\chi_{ac}^{\prime}$) and imaginary ($\chi_{ac}^{\prime\prime}$) parts of the ac susceptibility data in the $T$ range 10-88 K are depicted in fig.~\ref{acchi} (a) and (b) respectively. These measurements were performed in absence of any dc magnetic field, whereas the amplitude of the ac field was kept constant at 10 Oe. The anomalies near  $T_f$  and $T_s$ observed in dc magnetization data are also present in the ac measurements.  In $\chi_{ac}^{\prime}(T)$ data, $T_f$ is associated with the onset of  a sluggish drop, while $T_s$ is characterized by a change in slope. The corresponding anomalies are more prominent in the $\chi_{ac}^{\prime\prime}(T)$ data, where  broad peak-like structures are observed around $T_f$ and $T_s$. In the frequency ($f$) dependent ac measurement, $T_f$  shows strong variation with $f$. This indicates the onset of a spin freezing phenomenon (spin glass like state) or a blocking (in case of superparamagnetic clusters) around this anomaly. The shift in $T_f$  is found to be as high as 10 K for $f$ changing from 19 Hz to 191 Hz. Such large  shift  rules out the possibility of a canonical spin glass (CSG) state. The relative shift in $T_f$ can be expressed as $\mathcal{P}$ = $ \frac{\Delta T_f}{T_f [\Delta \log\omega]}$, where $\omega  = 2\pi f$ is the angular frequency of the ac excitation. $\mathcal{P}$ is typically found to be between 0.005-0.01 for CSG system where spin freezing at the atomic level takes place.~\cite{mydosh-book,schiffer}  For  cluster glass (randomly frozen interacting magnetic clusters of finite dimension), the value of $\mathcal{P}$ lies somewhat between 0.03 to 0.06. In case of  systems containing noninteracting superparamagnetic clusters, $\mathcal{P}$ is generally $\geq$ 0.1.~\cite{mydosh-book,kp-nio} Systems associated with complex spin freezing phenomena can also give rise to large value of $\mathcal{P}$ as observed in case of  disordered antiferromagnetic spin chain compounds,~\cite{sampath, hardy} or spin ice system.~\cite{snider} For the present sample, $\mathcal{P}$ was found to be 0.19 which may apparently indicate superparamagnetic like state in the alloy. The subsequent investigation shows a more complex spin freezing phenomenon in the system.  In $\chi_{ac}^{\prime\prime}(T)$, the anomaly associated with $T_s$ is quite prominent in the form of a peak. However, we failed to observe any shift in the peak position with varied $f$.

\par
 The $T_f$ versus $f$ data can be well fitted by the empirical Vogel-Fulcher (VF) law $f =  f_0 \exp [-E_a/k_B(T_f-T_{VF})]$.~\cite{mydosh-book, binder-rmp} Here $E_a$ is the activation energy of the spin glass (which determines the energy barrier for spins to align with the external magnetic field), $T_{VF}$ is the Vogel-Fulcher temperature, and $f_0$ is the characteristic frequency for spin freezing. Solid line (red) for the $f$ versus $T_f$ curve  in the inset of fig.~\ref{acchi}(b) indicates Vogel-Fulcher fitting to the experimental data. The fitted parameters, $E_a$, $f_0$, and $T_{VF}$, are found to be 45.8 K, 1.2$\times$10$^4$ Hz, and 36.8 K, respectively. The value of $f_0$  for the present sample  is found to be small compared to the reported values of CSG ($\sim$ 10$^{8}$ Hz).~\cite{mydosh-book} We failed to fit the $T_f$ versus $f$ data to a simple Arrhenius type law ($f = f_0 \exp [-E_a/k_BT_f]$) with physically meaningful value of $f_0$. Arrhenius type of behavior is generally expected for noninteracting superparamagnetic systems.

\par
We have  investigated the dynamical scaling (DS)  behavior of the frequency shift observed in $\chi_{ac}^{\prime\prime}$ data [see $\tau$ versus $T_f$ curve in the inset of fig.~\ref{acchi}(b)]. Relaxation time $\tau$ (for the decay of the fluctuations to the spin correlation length) can be expressed as $\tau = \tau_0(T_f/T_g-1)^{-z\nu}$~\cite{js}, where $z$ is the dynamic critical exponent, $\nu$ is the spin-correlation length exponent, $T_g$ is the zero frequency spin freezing temperature, and $\tau_0$ is the characteristic spin flipping time. Taking $\tau$ = 1/(2$\pi f$), the  parameters $\tau_0$, $T_g$ and $z\nu$, are found to be 7.5$\times$10$^{-4}$ s, 43.56 K and 1.63 respectively. For CSG, the value of $\tau_0$ was reported to be $\sim$10$^{-13}$ s.~\cite{mydosh-book}  Such  high values of $\tau_0$ obtained for the present sample indicate slower rate of spin flipping as observed  for several  cluster glass and reentrant spin-glass systems.~\cite{scott-prb,de-jap,tokura-prl,psanil-prb} In addition, the fitted value of $z\nu$ (=1.63)  is found to be considerably lower than values (4-12) reported for different SG systems.~\cite{mydosh-book} In recent times, low values of $z\nu$ were reported in few systems such as LaMn$_{0.5}$Fe$_{0.5}$O$_3$ and BiFeO$_3$.~\cite{de-jap,scott-prb} 

\begin{figure}[t]
\centering
\includegraphics[width = 8 cm]{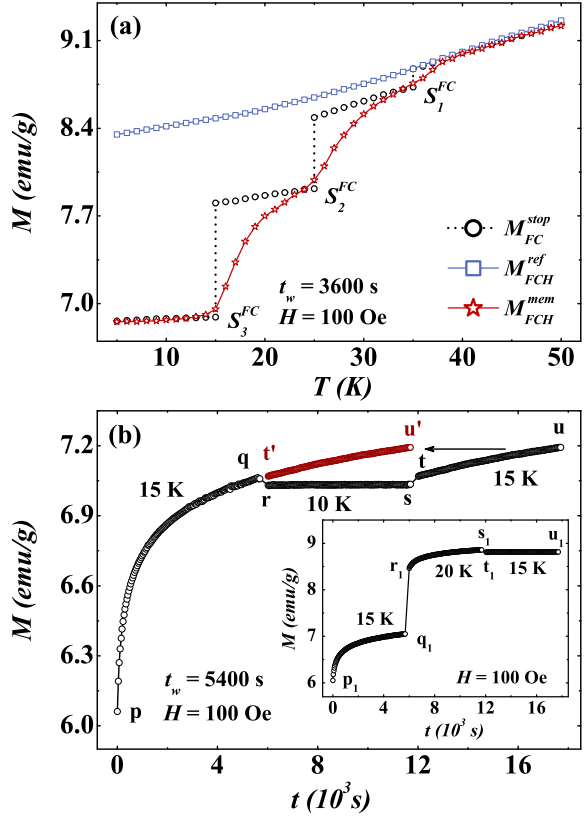}
\caption{(Color online) (a)  shows field cooled field stop  memory effect in dc magnetization versus temperature  data for Cu$_{2.836}$Mn$_{0.44}$Al$_{0.724}$ alloy. The memory measurement was performed by cooling the sample in $H$ = 100 Oe with intermediate zero-field stops at $T$ = 35, 25, and 15 K ($M^{stop}_{FC}$) followed by uninterrupted heating in 100 Oe ($M^{mem}_{FCH}$). The reference curve ($M^{ref}_{FCH}$) was measured on heating after the sample was field-cooled in 100 Oe without intermediate stops. (b) shows the time dependent magnetic memory measurement in presence of 100 Oe of applied field. The time variation of $M$ was measured in three consecutive segments {\it viz.}  $\overline{pq}$ (15 K), $\overline{rs}$ (10 K) and $\overline{tu}$ (15 K) with the duration of each segment being 5400 s. Here $\overline{t'u'}$ segment is obtained by simply shifting $\overline{tu}$ to merge its starting point with the end point of $\overline{pq}$. The inset of (b) shows the positive $T$ cycling or rejuvenation measurement of magnetic relaxation.} 
\label{fcmemory}
\end{figure}

\par
To shed more light on the low temperature glassy state of the sample, we recorded  $\chi_{ac}$ at different amplitudes of the applied ac magnetic field ($h_{ac}$) keeping the frequency of the ac signal constant at 97 Hz [see figs.~\ref{chiampl}(a) and (b)]. $h_{ac}$ has little  effect on $\chi_{ac}$ below $T_s$, however both real and imaginary $\chi_{ac}$ components  become highly sensitive to $h_{ac}$ above $T_s$. A large shift in $T_f$  ($\sim$ 10 K) for $h_{ac}$ changing from 4 to 8 Oe is observed in the $\chi_{ac}^{\prime\prime}(T)$ data. $h_{ac}$ dependence of $\chi_{ac}$ is not unexpected and it has been reported earlier in case of reentrant spin glass materials, phase separated manganites and magnetic superconductors.~\cite{nordblad-prb,psanil-prb2,awana-jap} Interestingly, for low amplitude of the ac excitation field ($h_{ac}$ = 1 Oe) no peak like feature around $T_f$ is observed in $\chi_{ac}^{\prime\prime}(T)$. The $h_{ac}$ dependence of $\chi_{ac}^{\prime\prime}(T)$ indicates strong nonlinearlity particularly around $T_f$.~\cite{hac} Below a certain threshold ac field, the frozen spins remain almost unperturbed by the ac excitation. The shift of $T_f$-peak with $h_{ac}$ is due to the competition between temperature and the driving ac field. The stronger the ac excitation, it can perturb the frozen spins down to lower temperature, and subsequently $T_f$ shifts to lower-$T$ with increasing $h_{ac}$.    

\begin{figure}[t]
\centering
\includegraphics[width = 8 cm]{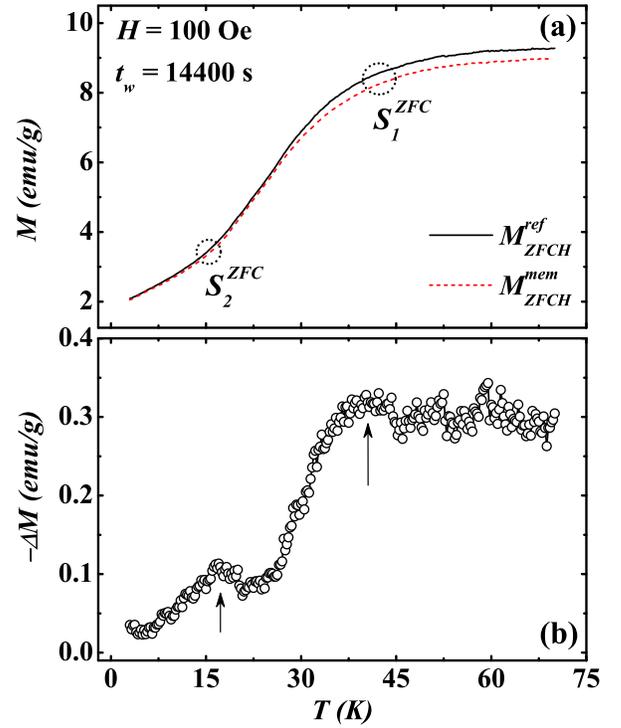}
\caption{(Color online) (a) shows the memory measurement in the zero field-cooled condition recorded on dc magnetization vs temperature data for Cu$_{2.836}$Mn$_{0.44}$Al$_{0.724}$ alloy. The sample was first cooled in $H$ = 0 down to 5 K, with intermediate stops at 42 and 15 K for 14400 s each. The sample was then reheated in $H$ = 100 Oe up to 70 K ($M^{mem}_{ZFCH}$). A zero-field cooled reference curve ($M^{ref}_{ZFCH}$) without intermediate stops during heating is  shown as a dotted line.  The difference in  magnetization $ \Delta M = M^{mem}_{ZFCH} - M^{ref}_{ZFCH}$ is plotted in (b).}
\label{zfcmemory}
\end{figure}

\par
The observed anomalies in $\chi_{ac}$ measurements indicate a frozen metastable state in the sample at least below $T_f$. Tempted by this observation, we performed  magnetic memory measurements (see figs.~\ref{fcmemory} and ~\ref{zfcmemory}) which can independently support the non-ergodic nature of the ground state as well as its probable origin. The field-cooled field-stop memory measurement (see fig.~\ref{fcmemory}(a)) in the $M$ versus $T$ data were recorded following the protocol described by Sun {\it et al.}~\cite{salamon-prl} Here the sample was cooled down to 5 K in 100 Oe with intermediate stops for $t_w$ = 3600 s at 35, 25 and 15 K. Subsequent heating in 100 Oe produces  characteristics wiggles at those selected temperatures confirming the presence of filed-cooled memory. No signature of memory was observed above $T_f$ indicating that the frozen non-ergodic state only exists below $T_f$. 

\par
Presence of magnetic memory in the alloy was  confirmed by the relaxation memory measurements with negative $T$ cycling (fig. ~\ref{fcmemory} (b)).~\cite{ab-prb,sc2-epl} Here relaxation data [$M$ versus time ($t$)] were recorded in 100 Oe of field at three consecutive segments $\overline{pq}$, $\overline{rs}$ and $\overline{tu}$ with the sample being held at 15 K, 10 K and 15 K respectively for 5400 s each. The segment $\overline{tu}$ shows a continuation of $\overline{pq}$. This memory effect reflects that the state of the system before cooling is recovered when the sample is cycled back to the initial $T$. To confirm the reliability of the observed memory, we  performed positive $T$ cycling or rejuvenation, where the sample was heated to a higher value of $T$ for the intermediate relaxation.~\cite{sg-prb,ab-prb} Such rejuvenation measurement has been depicted in the inset of ~\ref{fcmemory}(b). Here $M$ versus $t$ was measured consecutively  at 15 K ($\overline{p_1q_1}$), 20 K ($\overline{r_1s_1}$) and 15 K($\overline{t_1u_1}$). Clearly, $\overline{t_1u_1}$ does not follow the trend of $\overline{p_1q_1}$ signifying that the magnetic state is lost on intermediate heating. This is a typical signature of glassy or superparamagnetic state below the freezing or blocking temperature.   

\par
It is now essential to distinguish between superparamagnetism and a spin-glass like ground state for the studied sample. The Vogel-Fulcher like behavior of the $f$-$T_f$ variation in $\chi_{ac}$ indicates the existence of finite magnetic interaction among spins and/or spin clusters resulting a spin-glass like state below $T_f$.  To confirm this, we performed zero-field-cooled memory measurements in the $M$ versus $T$ data (see fig.~\ref{zfcmemory}).~\cite{ab-prb,sc2-epl,sg-prb} Here sample was cooled from 300 to 5 K  in $H$ = 0 with intermediate stops for 14400 s at 42 K and 15 K (points $S^{ZFC}_1$, and $S^{ZFC}_2$ respectively in fig.~\ref{zfcmemory}(a)) After that the sample was heated back to 70 K in presence of $H$ = 100 Oe. Substantially strong signature of anomalies are observed on the difference curve $ \Delta M (T) = M^{mem}_{ZFCH}(T) - M^{ref}_{ZFCH}(T)$ at the stopping temperatures providing positive signature of memory [fig.~\ref{zfcmemory}(b)]. This is a clear indication that the ground state of the system is spin glass like.

\section{Discussion}
Our investigation indicates that the studied alloy Cu$_{2.84}$Mn$_{0.44}$Al$_{0.72}$ undergoes long range FM ordering below $T_C$ = 270 K and it attains a glassy magnetic state below $T_f$ = 65 K. The magnetic state between $T_C$ and $T_f$ is certainly FM as evident from finite coercivity and spontaneous magnetization rather than a superparamagnetic state claimed previously for certain Cu-Mn-Al alloys.~\cite{yef,Pardo-am, obrado-sg} Development of a glassy magnetic state out of a long range magnetically ordered state is referred as a reentrant spin glass (RSG) transition.~\cite{rsg-prl, rsg-epl, rsg-dho, rsg-prb} RSG state occurs due to the competition between FM and AFM interactions arising from site or bond randomness~\cite{rsg-prb}, although the mean interaction has to have a nonzero FM character for the realization of an RSG phase. The magnetic phase below the spin freezing transition (here occurring at $\sim$ $T_f$) is a mixed phase with the coexistence of FM and spin-glass orderings. In contrast, usual CSG have spin freezing directly from a disordered paramagnetic phase and it is generally not associated with any net FM moment. The characteristic parameters (such as $\mathcal{P}$, $\omega_0$,  $\tau_0$ etc.) obtained from $f$ dependent $\chi_{ac}$ measurement and their subsequent fitting to the relevant models do not  match with that of CSG. They are rather close to the values reported for cluster glass or RSG systems.      

\par
The magnetic phase in Cu$_{2.84}$Mn$_{0.44}$Al$_{0.72}$ is substantially different from the conventional RSG systems such as Fe-Au or Fe-Mn alloys. The glassy phase in conventional RSG develops from an ordered FM phase, whereas in the present alloy the spin freezing is preceded by a first order structural transition which itself  modifies the high-$T$ FM phase. In case of Cu-Mn-Al alloys the sign of the magnetic interaction depend on the Mn-Mn distance (due to the RKKY mechanism). For the stoichiometric Heusler compound Cu$_2$MnAl with ordered L2$_1$ structure, Mn-Mn distance is next nearest neighbor (nnn) type and this gives rise to FM interaction. For the present sample, some excess Mn is doped randomly at the Al and Cu sites. This will reduce the Mn-Mn distance in some sites paving the path for random AFM type bonds in the backdrop of FM correlations. It has been argued by Prado {\it et al.}~\cite{prado-cumnal} that percentage of Mn-Mn AFM bonds enhances in presence of chemical disorder  below the MT due to the increased nearest neighbor Mn-Mn occupancy. The occurred  glassy state in Cu$_{2.84}$Mn$_{0.44}$Al$_{0.72}$ just below the MT is supposedly connected to the structural transition induced random AFM bond formation.

\par
The most intriguing result of the present study is the anomalies observed in the $\chi_{ac}$ data on and around the spin freezing temperature. In $\chi_{ac}^{\prime\prime}$ versus $T$ data, two distinct features were seen, one at $T_f$ and another at slightly low temperature denoted by $T_s$. As evident from our previous discussion, $T_f$ denotes a spin freezing point characterized by large frequency shift in ac $\chi$ measurement. This is  approximately the onset point of thermomagnetic irreversibility between ZFCH and FCH curve in $M$ versus $T$ measurements. The irrevrsibility starts to grow rapidly only when the sample is cooled approximately below $T_s$.  $H_{coer}(T)$ [see fig.~\ref{mtmh}(b)] also contains a signature of $T_s$ in the form of a minimum. 

\par  
There are several theoretical works related to RSG state, however still now the origin and nature of such phase is debated. The two major theoretical models which describes the RSG state are (i) mean field  treatments of Heisenberg spins with infinitely long-range interaction~\cite{rsg-GT1, rsg-GT2, mydosh-book, binder-rmp} commonly known as Gabay-Toulous (GT) model and (ii) a phenomenological random field model.~\cite{rsg-aeppli} In the mean field model it has been argued that spin freezing in RSG phase in presence of a magnetic field takes place through set of transitions lines in the $J_0$-$T$ phase diagram, where $J_0$ is the exchange interaction averaged over all the magnetic bonds. It was predicted that there exist three characteristics temperatures $T_C$, $T_{GT}$ and $T_{AT}$. Below $T_C$, the FM phase develops from the paramagnetic state. Below $T_{GT}$, the system enters into a mixed phase $\mathcal{M}_1$, where transverse component of the spins (with respect to the direction of $H$) freezes keeping the longitudinal components ferromagnetically ordered. On further cooling below $T_{AT}$, a crossover to a second mixed phase ($\mathcal{M}_2$) occurs which is associated with strong magnetic irreversibility in the longitudinal component of the spins. Experimental support for such model is found in several metallic spin glass systems.~\cite{rsg-kim, rsg-prb} It is to be noted that $T_{AT}$ is a crossover temperature rather than a transition below which non-ergodicity sets in gradually in the longitudinal spin components.

\par
For the present sample, the observation of two  anomalies presumably indicate an RSG phase similar to the prediction of mean field Heisenberg model. In that situation $T_f$ and $T_s$ of our sample correspond to $T_{GT}$ and $T_{AT}$ respectively. The large peak-shift in $\chi_{ac}(T)$ with $f$ and $h_{ac}$ and the onset of irreversibility between ZFCH-FCH magnetization at or around $T_f$ certainly signifies the spin freezing. On the other hand $T_{s}$ shows almost negligible peak shift with $f$ or $h_{ac}$ along with strong irreversibility in $M$ mimicking the properties of the crossover point $T_{AT}$. Our memory measurements (both in field cooled and zero-field-cooled protocols) show positive signature below $T_f$ as well as below $T_s$. Therefore the glassy state that develops below $T_f$ continues to exist at the lowest temperature. 

\par
The exponent $z\nu$ obtained from the dynamical scaling analysis of frequency shift in $\chi_{ac}$ data is quite low for the present alloy. The spin freezing temperature $T_f$ of Cu$_{2.84}$Mn$_{0.44}$Al$_{0.72}$ actually falls within the region of thermal hysteresis associated with MT. A possible mechanism related to domain wall dynamics was mooted for the observed low value of $z\nu$ in BiFeO$_3$.~\cite{scott-prb} The MT give rise to low temperature martensite with structural variants, which can actually have similar effect on the dynamics of spin freezing.      

\par
In conclusion, we present a comprehensive view of the glassy magnetic state of Cu$_{2.84}$Mn$_{0.44}$Al$_{0.72}$ alloy.  The glassy state is reentrant type and it is likely to be connected to the first order structural phase transition observed in this ferromagnetic shape memory alloy. This is an unique example where an RSG phase arises from the result of a martensitic type first order structural transition showing  GT like character of the reentry. It is to be noted that previous reports on Cu-Mn-Al alloys by Obrad\'o {\it et al.}~\cite{obrado-sg} indicated a superparamagnetic ground state  for large Mn concentration. The presently studied sample have different Cu:Mn:Al ratio, it shows a  spin-glass like state even for large concentration of Mn. This indicates that the glassy state in Cu-Mn-Al alloys may depend on factors other than Mn concentration. 
\par
S. Chattopadhyay wishes to thank CSIR (India) for his research support. The present work is supported by the grants from BRNS, India (project no. 2012/37P/39/BRNS). Authors would also like to thank Department of Science and Technology, India for low temperature and high magnetic field facilities at UGC-DAE Consortium for Scientific Research, Kolkata center (project number:IR/S2/PU-06/2006).


%

\end{document}